# LA NORME TECHNIQUE COMME CATALYSEUR DE TRANSFERT DE CONNAISSANCES : LA FRANCOPHONIE A L'ŒUVRE DANS LE DOMAINE DE L'EDUCATION

Mokhtar Ben Henda,
ISIC, Université Bordeaux Montaigne, France

## Entre l'humain et la technique, tout est objet de normes

La normalisation est un principe régulateur ancré dans l'histoire humaine depuis l'Antiquité pour soutenir les échanges et le développement des sociétés. L'histoire nous montre à ce propos que les mécanismes de normalisation ont toujours joué un rôle fondamental dans l'organisation des sociétés. Au Moyen-Âge, par exemple, les corporations de métiers imposaient des standards pour régir la qualité des produits, la production et le commerce. Les normes, bien ancrées dans les lois et règles coutumières, garantissaient le bon déroulement des transactions, des mariages et des successions, instaurant ainsi un système normatif accepté par la société (Dupuis-Déri, 2017, p. 141). L'invention de l'imprimerie par Gutenberg vers 1450, inspirée de son savoir-faire en orfèvrerie, est un bon exemple de l'importance de la normalisation des caractères mobiles pour la production et la diffusion en masse des connaissances. Au cours des périodes suivantes, les normes ont facilité l'essor économique notamment dans les villes dans lesquelles les systèmes de mesures et de poids étaient essentiels au bon fonctionnement des échanges commerciaux (Beaulande-Barraud et Marmursztejn, 2016; Gauvard et al., 2003; Jimenez-Garrido, 2013; White, 2021). Avec la Révolution industrielle, la normalisation a pris une dimension institutionnelle permettant que les procédés de fabrication engendrent une production homogène à grande échelle.

L'industrie ferroviaire, par exemple, a imposé des normes concernant l'écartement des rails et la synchronisation des horaires pour garantir la fiabilité des transports. L'apparition du télégraphe a introduit ses propres normes de communication, comme le code Morse, tandis que l'urbanisation et la construction de grands édifices étaient facilitées par la production de poutrelles métalliques standardisées.

Notre époque contemporaine n'est, à ce titre, que le prolongement d'une longue histoire des techniques et d'une lente évolution d'un processus normatif complexe. De nos jours, ère de la mondialisation, il n'est pratiquement pas un seul des grands secteurs d'activité qui ne soit entièrement ou partiellement encadré par des normes ou des standards. Selon l'historien Olivier Borraz, sociologue au CNRS : « L'extension de la normalisation est indissociable, bien qu'elle les précède souvent, des phénomènes de globalisation économique et de transformation des processus de régulation politique à l'échelle internationale, régionale, voire nationale » (Borraz, 2005, p. 123).

C'est une preuve de résilience de ce concept qui prend aujourd'hui une forte connotation numérique dans un monde envahi d'une infinité de produits technologiques, services et ressources numériques dont nous ne comprenons pas très souvent les modes de fonctionnement. Pour Bertrand Gille, historien des techniques :

> En règle générale, toutes les techniques sont, à des degrés divers, dépendantes les unes des autres et il faut nécessairement entre elles une certaine cohérence : cet ensemble de cohérences aux différents niveaux de toutes les structures, de tous les ensembles et de toutes les filières compose ce qu'on peut appeler un système technique. (Gille, 1978, p. 19).

La cohérence d'un système technique implique souvent un système social correspondant. Dans son ouvrage *le mode d'existence des objets techniques*, publié pour la première fois en 1958, Gilbert Simondon en fait l'analyse comme une influence qu'exerce la technique sur la culture et les pratiques sociales et la manière comment elle s'intègre et évolue au sein des sociétés entre rejet et acceptation, loisir et travail, théorie et pratique. Pour Simondon, l'objet technique a toujours été au cœur de cette dialectique et se définit essentiellement par son caractère utilitaire, mais aussi par son impact cognitif et socioculturel pour être accepté ou rejeté :

> La condition première d'incorporation des objets techniques à la culture serait que l'homme ne soit ni inférieur ni supérieur aux objets techniques, qu'il puisse les aborder et apprendre à les connaître en entretenant avec eux une relation d'égalité, de réciprocité d'échanges : une relation sociale en quelque manière. (Simondon, 1989, p. 88).

Ce que l'on peut déduire à ce propos, c'est que dans notre civilisation contemporaine, un décalage (ou un « hiatus » comme l'appelle Simondon) caractérise d'une part la manière comment l'homme réagit face aux objets techniques et de l'autre, ce que ces objets sont réellement. Il s'agit d'un lien souvent flou qui engendre chez l'homme une série de jugements empreints de mythes, mêlant exagération et dépréciation. Pour dépasser cette vision confuse, il est essentiel de prendre conscience du véritable « mode d'existence des objets techniques » et de reconstruire une relation fondée sur la compréhension plutôt que sur l'illusion.

Ce processus se perpétue actuellement avec l'avènement de l'intelligence artificielle (IA) qui suscite des débats ontologiques profonds entre l'humanisme et sa valorisation des capacités morales et intellectuelles de l'homme, le transhumanisme qui prône l'augmentation de l'humain grâce à la technologie et le post-humanisme dans sa quête d'une perspective fusionnelle entre l'homme et la machine. Or, comme le souligne Pierre Joliot, professeur au Collège de France, ces oppositions ne sont pas toujours anticipées, car « aucune des révolutions successives qui ont marqué le monde depuis le début de l'ère industrielle n'a été prévue par les futurologues qui en sont toujours réduits à des extrapolations hasardeuses » (Joliot, 2001, p. 38).

Dans cette maïeutique sociotechnique, il faut également souligner que dans l'évolution de la société humaine, les changements ont souvent lieu selon deux principes antinomiques. Alors que le premier prévoit que la « résistance au changement » reste un phénomène intrinsèque à la nature humaine, le deuxième stipule, en revanche, que « nul ne peut arrêter le progrès », un vieil aphorisme ou adage signifiant tacitement que les avancées technologiques et sociales sont inévitables et irréversibles. Nous en sommes aujourd'hui témoins avec, toutefois, une différence significative : l'innovation ne correspond plus à l'adage d'antan « le besoin crée l'invention », mais traduit plutôt une forme d'aliénation d'une société consumériste, amplifiée par la concurrence et le battage médiatique. On le voit à travers l'abondance et la redondance de certains produits comme la multiplication ludique des réseaux sociaux, voire de certains produits futiles et puériles comme l'invention de parapluies ou des matelas intelligents ! Nous sommes désormais dans une société de consommation que Frédéric Beigbeder, écrivain et critique littéraire français, qualifie de « Société de Tentation » (Beigbeder, 2007, p. 10). Pierre Joliot le résume à son tour : « Une société qui survit en créant des besoins artificiels pour produire efficacement des biens de consommation inutiles ne paraît pas susceptible de répondre à long terme aux défis posés par la dégradation de notre environnement » (Joliot, 2001, p. 140).

## Complexité d'une notion polysémique

À notre avis, l'idée que tout doit être normalisé pour être fonctionnel repose à l'origine sur un concept fondamental des théories des systèmes et de l'organisation sociotechnique de l'environnement humain. S'il est admis que la normalisation favorise la compatibilité et l'interopérabilité et réduit la complexité en particulier dans les systèmes sociaux, économiques et technologiques complexes, ce jugement arbitraire soulève des controverses autour des principes opposés de flexibilité et d'uniformité, de souplesse et de rigueur, de créativité et de formatage, d'individualité et de communautarisme. Les divergences qui caractérisent chacune de ces valeurs antinomiques en font des sujets de controverses. Elles nécessitent dès lors quelques précisions du champ sémantique de l'acte normatif, d'autant qu'il en existe plusieurs. La plus courante est sans doute entre « norme » et « standard » dans la langue française, car en anglais, même si le terme « *standard* » domine largement celui de « *norm* », les deux sont plus souvent compris dans un sens large relatif à l'interaction sociale (Burton, 1978, p. 1). Il existe en effet un flou entre les langues anglaises et françaises d'autant que « le terme anglais de standard, désigne des réalités qui, en français, sont susceptibles d'être traduites de trois façons spécifiques : le standard, la norme et l'étalon (au sens de l'étalon de mesure) » (Lelong & Mallard 2000 : 10). On différencie également la norme *de jure* (ou « de droit »), qui correspond à un document de référence approuvé par un institut de normalisation reconnu, de la norme *de facto* (ou « de fait »), qui correspond plutôt à ce que l'on appelle « standard » en français (Manceau, 2012). Brunsson et Jacobsson (1998), rapportés par Martinsson, élargissent cette ambiguïté en y intégrant les directives comme une variante normative :

> Les directives sont explicites et émises par une personne ayant autorité ou pouvoir, comme un législateur ou la haute direction d'une organisation. Les normes, en revanche, ne sont pas nécessairement écrites. Les normes sont des règles intériorisées qui peuvent être suivies sans être réfléchies. (Martinsson, 2009, p. 29).

Sauf que cette acception pose plus d'un problème quand elle est confrontée à ce que d'autres affirment, comme Mark Banik, chercheur à l'Université du Québec :

> Une norme est un référentiel, une spécification technique publiée par une association de producteurs ou une organisation de normalisation, comme l'Organisation internationale de normalisation (ISO) […], un standard ne fait pas l'objet d'un référentiel publié, mais plutôt d'une convention adoptée par une bonne proportion d'utilisateurs ou de producteurs. (Banik, 2015, p. 158).

La notion de norme véhicule ainsi une ambiguïté sémantique à plusieurs niveaux et désigne bel et bien une réalité hétérogène et difficile à cerner dans laquelle la dimension technique l'emporte largement. Il est dès lors naturel de se poser la question sur le lien potentiel pouvant exister entre normes techniques et normes sociales dans la complexité du monde réel.

### La norme technique comme vecteur de cohésion socioculturelle et linguistique

Si notre société contemporaine évolue à pas très sûrs vers les systèmes intelligents et autonomes grâce à l'émergence de l'intelligence artificielle très axée, elle-aussi, sur des normes techniques, de nombreuses organisations comme l'Organisation Internationale de Normalisation (ISO), la *International Electrotechnical Commission* (IEC), l'Union Internationale des Télécommunications (UIT), le Consortium du *Word Wide Web* (W3C), l'*Internet Engineering Task Force* (IETF), etc. élaborent en permanence des normes sur la base de consensus internationaux en vue d'assurer une utilisation éthique, durable et responsable des technologies émergentes. Tant que le monde devient de plus en plus interconnecté, les normes et standards techniques aident à gérer la complexité inhérente aux systèmes globaux, renforçant leur résilience et leur réactivité. Ces normes et standards, outre qu'ils sont de nature technique, sont établis par une combinaison de facteurs culturels, sociaux et historiques et servent de lignes directrices pour le comportement et les interactions des individus au sein de la société mondiale. À titre d'exemple, plusieurs normes de l'*Institute of Electrical and Electronics Engineers* (IEEE), organisme pourtant dédié à l'avancement de la technologie, intègrent les politiques et principes mondiaux des droits de l'homme, de la diversité culturelle et linguistique, des préoccupations environnementales et du bien-être sociétal. La norme IEEE 1680-2009, par exemple, destinée à la réglementation de l'impact des produits électroniques sur l'environnement, ou encore la norme IEC 63318 pour l'accès à l'électricité rurale via les micro-réseaux ne sont que des illustrations de l'entière adhésion des organismes de normalisation internationaux aux 17 objectifs de développement durable des Nations-unies. Les normes techniques agissent ainsi comme des catalyseurs de convergence et des vecteurs de communication entre différentes langues et cultures en créant un cadre de référence partagé qui transcende les particularités locales et régionales.

De fait, les normes techniques servent de fondement à une infrastructure technologique globalisée, favorisant un accès équitable aux connaissances, aux savoirs et savoir-faire. Bien qu'elles assurent que les *smartphones* et tablettes connectées en WiFi (norme IEEE 802.11) puissent être utilisées partout dans le monde, malgré les différences dans leur fabrication, les normes techniques ne se limitent pas à des prescriptions fonctionnelles de processus technologiques, mais contribuent aussi à la structuration d'une culture mondiale, où les acteurs issus de divers horizons peuvent communiquer et coopérer de manière transparente, fluide et efficiente. C'est le cas aussi des normes de l'ISO, de l'IEC, de l'UIT, et d'autres structures internationales de normalisation, dont les comités techniques sont souvent composés de représentants de divers pays et diverses cultures.

Ces comités cherchent à établir des consensus favorisant l'interopérabilité tout en respectant les particularités régionales et locales. En effet, les besoins en normes peuvent varier en fonction des contextes socioculturels, des spécificités linguistiques et des traditions distinctes au sein même de différentes communautés. Ces variations peuvent être influencées par les usages linguistiques — comme les styles de parole, la grammaire et le vocabulaire — dans les communications et l'éducation, mais aussi par les dynamiques de pouvoir, les inégalités sociales et les héritages historiques. C'est aussi face aux problèmes d'accessibilité et d'inclusion, de données personnelles, de sécurité et de vie privée que la normalisation joue un rôle déterminant (e.g. la norme *Web Content Accessibility Guidelines* (WCAG) du W3C, l'ISO 22739 de la *Blockchain* et l'ISO 20022 de la cryptomonnaie).

## La normalisation comme vecteur culturel de transfert de connaissances dans le domaine de l'éducation

Depuis 1962, date de création de l'*American National Standards Institute* (ANSI), les normes techniques jouent un rôle central dans l'éducation. Elles ont comme objectif l'harmonisation des pratiques et standards éducatifs. Avec la création de l'*Aviation Industry CBT Committee* (AICC) en 1988, a commencé l'usage à grande échelle des normes et des spécifications pour les produits d'enseignement à distance (Blandin, 2003, p. 3).

Ce virage a marqué un changement majeur : alors que la normalisation traditionnelle reposait sur un consensus technique assurant la qualité des objets physiques et des procédés industriels concrets, la normalisation des technologies de l'information et de la communication en éducation vise désormais des représentations numériques et des opérations abstraites.

Cette différence pose des défis majeurs, car au-delà d'un simple accord technique, ces représentations touchent à des conventions sociales et pratiques culturelles (Blandin, 2003, p. 1). Pour l'illustrer, les normes industrielles, comme la norme ISO 9001 (Systèmes de management de la qualité), centrées sur la qualité des produits et procédés concrets, visent des résultats mesurables et reproductibles. En revanche, les normes éducatives, telles que l'ISO/IEC TR 29163 (mieux connue norme SCORM (*Sharable Content Object Reference Model*) pour l'interopérabilité des contenus numériques, la norme WCAG (ISO/IEC 40500) pour l'accessibilité numérique, ou la norme ISO/IEC 27701 pour le management des informations sur la vie privée, etc. portent sur des aspects plus abstraits influencés par des contextes sociaux et culturels, impliquant un large éventail de parties prenantes : enseignants, chercheurs, apprenants. La normalisation des langues et des modèles de communication linguistique, la normalisation des vocabulaires et des ontologies, des lexiques et des traductions comme celles définies par le Sous-comité 37 de l'ISO, renforcent également la capacité de comprendre, d'exprimer et de communiquer efficacement au-delà de la diversité des langues. Ces normes facilitent le transfert de connaissances et contribuent à la normalisation des curriculums, des qualifications académiques et des critères d'évaluation, permettant aux acteurs éducatifs de s'aligner sur des objectifs communs.

Comme nous le soulignons plus loin, la normalisation en éducation fait référence à l'appropriation cohérente des technologies émergentes pour harmoniser l'écosystème mondial de l'éducation dans ses divers aspects systémiques dont l'interopérabilité des ressources pédagogiques, la mobilité étudiante, la terminologie et la sémantique pédagogiques, l'équité et l'accessibilité numérique à l'éducation, l'assurance qualité, les données personnelles et la vie privée, etc. L'adoption et l'adaptation périodique des normes permettent aux institutions éducatives de partager efficacement des ressources et des pratiques pédagogiques, facilitant l'accessibilité des savoirs à l'échelle mondiale. Ces normes structurent et réduisent les disparités en matière de qualité et de format des ressources, soutenant la mutualisation des expertises et des savoir-faire. Elles favorisent également l'alignement entre les pratiques éducatives et les technologies, créant un environnement propice à l'innovation et à l'adaptation locale. C'est le cas propre au sous-comité 36 du premier comité technique joint (JTC1) entre l'ISO et l'IEC (ISO/IEC JTC1 SC36), dont les termes de références soulignent qu'il est chargé de la « normalisation dans le domaine des technologies de l'information pour l'apprentissage, l'éducation et la formation à destination des personnes, groupes ou organismes, et en vue de permettre l'interopérabilité et la réutilisation des ressources et des outils » (ISO, 2020).

La Francophonie, par l'intermédiaire de l'Agence universitaire de la Francophonie (AUF), a été un acteur clé du SC36 dès sa réunion fondatrice de Londres en 1999. Elle y bénéficie d'un statut de liaison de rang A. Bien qu'elle n'ait pas de droit de vote, l'AUF joue un rôle de recommandation essentiel, témoignant de l'importance stratégique de cette normalisation pour l'éducation et l'apprentissage dans l'espace francophone.

### Le cadre normatif des technologies éducatives au sein de l'ISO/IEC JTC1 SC36

Le sous-comité ISO/IEC JTC1 SC36 a été créé lors de la réunion plénière du JTC1 en 1999 à Séoul avec pour objectif de définir un modèle de référence mondial permettant la réutilisation et le partage de ressources pédagogiques numériques. La première réunion plénière du nouveau sous-comité s'est tenue en 2000 à Londres en présence d'une délégation de l'AUF. En 2003, l'AUF accueille la 7$^e$ plénière au Château de Versailles et inaugure en 2005 le premier *Open forum* scientifique de normalisation en marge du 2$^e$ Somment mondiale de la société de l'information (SMSI) en Tunisie.

La mission du SC36 est de créer des normes pour le marché des technologies éducatives en adoptant et en adhérant aux principes essentiels d'ouverture mondiale, de transparence, de cohérence technique et de consensus entre tous les pays membres (*P-Members*) et les pays observateurs (*O-members*) ainsi que les organisations de liaison (LO) comme l'AUF.

### Structure et organisation

Dans sa structure actuelle, le SC36 est constitué de groupes de travail (*Working Groups*) qui effectuent des tâches spécifiques décrites dans leurs termes de référence. Parmi les groupes durables (puisque certains sont dissous dès qu'ils n'ont plus de projets de normes en cours), on compte notamment ceux qui normalisent le vocabulaire (WG1 puis TCG), les profils apprenants (WG3), les métadonnées pédagogiques (WG4), la diversité culturelle et linguistique et l'accessibilité numérique (WG7), l'analyse des données d'apprentissage (WG8) et la modélisation des cours en ligne (WG9).

Le SC36 évolue aussi en fonction des technologies émergentes à impact éducatif auxquelles il crée des groupes *ad hoc* provisoires pour traiter des sujets d'actualité comme les *blockchains*, l'intelligence artificielle, les jumeaux numériques, etc. Ces groupes *ad hoc* sont dissous une fois qu'ils produisent, par consensus de toutes les parties prenantes, et publient, par un vote majoritaire, des rapports techniques ou des normes internationales.

Le SC36 fonctionne aussi en étroite collaboration avec d'autres instances de normalisation (Comités techniques/TC et sous-comités/SC), à la fois internes et externes à l'ISO, à l'IEC et au JTC1, afin d'éviter tout travail contradictoire ou redondant. En interne, des liaisons sont mises en place avec des instances travaillant sur des thématiques connexes comme la « Gestion et échange de données » (SC32), les « Interfaces utilisateur » (SC35), la « Terminologie et autres ressources linguistiques et de contenu » (TC37), la « Gestion de la qualité et assurance qualité » (TC39), les « Services d'apprentissage en dehors de l'éducation formelle » (TC232), etc. Parmi les liaisons externes à l'ISO ou à l'IEC, les plus actives sont celles avec l'Agence universitaire de la Francophonie (AUF), l'*International Information Centre for Terminology* (Infoterm), le *Learning Technology Standards Committee* de l'IEEE (LTSC) et le *EdTech Consortium*.

Le processus d'élaboration des normes du SC36, aligné sur les pratiques codifiées de l'ISO, débute par la soumission d'un projet répondant à un besoin identifié dans un secteur particulier. Un organisme membre de l'ISO, tel qu'un comité technique (TC), un sous-comité (SC), une délégation nationale (NB), une organisation de liaison (LO) ou un expert, propose un projet qui sera soumis au vote des pays membres permanents. Le projet est ensuite partagé pour être commenté et discuté avant que plusieurs cycles de vote permettent d'atteindre un consensus large comme condition de sa publication comme norme internationale. En général, le cycle complet de création d'une norme, depuis sa proposition initiale comme projet (*New Work Item Project*/NWIP) jusqu'à sa publication comme norme internationale (IS), dure environ 3 ans.

De fait, le processus de normalisation comporte plusieurs étapes, outre la procédure de « vote express » ou *fast track* qui consiste à ce qu'un projet, avec un certain degré de maturité ou la révision d'une norme déjà publiée, soit soumis directement à l'approbation comme projet final de norme internationale (IS) sans passer par les phases régulières de publication d'une norme. Le processus régulier est comme suit :

1. Stade de proposition : le besoin d'une nouvelle norme internationale dans le domaine est exprimée par un expert, un organisme ou une délégation nationale qui propose de voter au sein du comité d'une nouvelle étude de projet (*NWIP*) ;

2. Stade de préparation : un groupe de travail (WG) est généralement constitué par le comité responsable dans le but de préparer le projet de travail (*Working Draft/WD*). Le groupe de travail se compose d'experts et d'un animateur (généralement le chef de projet) ;

3. Stade de comité : à ce stade non obligatoire, des versions successives du *WD* sont diffusées en vue d'atteindre un consensus sur une version optimisée qui pourra passer à l'étape suivante de Document de Comité (*Committee Draft/CD*) ;

4. Stade d'enquête : le *CD* est transmis à l'ensemble des membres de l'ISO pour subir un vote et faire l'objet de commentaires sur une période de 12 semaines. Il est approuvé si deux tiers des membres participants votent en sa faveur et si les votes négatifs ne dépassent pas un quart du total. En cas d'approbation sans modification technique, le projet passe directement à la publication en tant que norme internationale ; sinon, il doit passer par l'étape d'approbation finale (*Final Draft International Standard*/FDIS) ;

5. Stade *FDIS* : lorsque ce stade est nécessaire, le projet est diffusé à tous les membres pour un vote de 8 semaines. Le projet est adopté comme norme internationale (IS) si deux tiers des membres participants votent en sa faveur et si les votes négatifs ne dépassent pas un quart du total ;
6. Stade de publication (*IS*) : le texte final est soumis pour publication. Seules des modifications rédactionnelles mineures sont apportées avant la publication officielle de la Norme Internationale.

En 2024, l'ISO/IEC JTC 1/SC36 compte 56 normes publiées et 13 projets en cours qui doivent passer les six étapes indiquées avant la publication comme norme internationale. Les 23 pays membres et les 26 pays observateurs sont représentés par des délégations nationales mandatées par les organismes nationaux de normalisation comme l'Afnor pour la France, la DIN pour l'Allemagne ou le KATS pour la Corée.

Contributions francophones

Par la contribution de l'AUF au SC36, la Francophonie joue un rôle actif dans le développement et la promotion des normes éducatives à l'échelle internationale. Elle mise particulièrement sur l'axe de la diversité culturelle et linguistique. Pour y parvenir, l'AUF met en œuvre une coopération stratégique avec l'Afnor (France) et le soutien de délégations de pays francophones comme le Conseil canadien des normes (CCN). Au moment de son adhésion au SC36, l'AUF partait de la conviction que les normes techniques facilitent l'accès à des ressources partagées et à des outils numériques dans les langues locales partenaires, renforçant ainsi l'appropriation culturelle des technologies et favorisant une inclusion numérique croissante. En intégrant les perspectives linguistiques et culturelles dans la normalisation, l'AUF envisage soutenir une cohérence socioculturelle durable qui transcende les frontières tout en respectant les identités régionales et locales.

Sa contribution à la normalisation des technologies éducatives a rapidement pris forme de prise de position dans la gouvernance du SC36. Avec la collaboration de l'Afnor comme représentant officiel de la France, habilité à occuper des rôles de responsabilité, l'AUF a pris depuis 2009 la coordination du groupe de travail sur le vocabulaire (WG1) alors que l'Afnor coordonne celui chargé de l'information sur les apprenants (WG3) et contribue fortement à celui sur les métadonnées pédagogiques (WG4).

L'une des contributions francophones majeures est la norme ISO/IEC 19788, connue sous l'acronyme MLR (*Metadata for Learning Resources*), qui répond au besoin de pallier les limites de la norme LOM (*Learning Object Metadata*), publié initialement en 2002 par l'IEEE et devenu difficile à gérer en raison de sa complexité et de ses profils d'application multiples et variés.

LOM est un schéma de métadonnées complexe, compte tenu du grand nombre d'éléments de description qu'il propose et du peu de structures aptes à employer intégralement. Largement répandu sous forme de « profils d'application » (modèles adaptés pour une application contextualisée tout en gardant l'interopérabilité avec le schéma source), LOM fait l'unanimité auprès de beaucoup d'organismes producteurs de ressources pédagogiques et d'environnements d'apprentissage numérique. En France, il a donné lieu aux profils d'application national LOM.fr qui, à son tour, a été décliné en ScoLOM.fr pour les ressources de l'enseignement secondaire et SupLOM.fr pour celles de l'enseignement supérieur. Le mouvement du libre accès (l'éducation ouverte en particulier) en a fait l'outil principal pour « normaliser » la description des ressources éducatives libres (REL). Toutefois, au fil du temps, les profils d'application du LOM ont fini par se multiplier au point de s'éloigner substantiellement du schéma source, perdant ainsi une grande marge d'interopérabilité entre eux.

Pour freiner la dérive du LOM, l'ISO a proposé, via le SC36, d'initier le projet de la norme MLR sur la base des spécifications de la norme ISO 11179 pour les « Registres de métadonnées ». L'atout majeur de cette norme est qu'elle permet de dissocier l'aspect conceptuel de l'aspect représentation lors de la conception de schéma de métadonnées, favorisant ainsi une compréhension commune de leur sémantique. Comme le souligne Yolaine Bourda :

> « Dans le LOM, le nom d'un champ est confondu avec sa signification, or celle-ci n'est pas neutre : elle dépend du contexte, de la culture, de la langue. L'idée est donc de s'intéresser aux définitions à un niveau abstrait, afin que tous, quels que soient leur culture et leur langue, puissent s'accorder sur la signification de chaque élément, sans se focaliser sur son nom. » (Bourda, 2004, p. 3).

Cette nouvelle norme doit donc respecter les investissements précédents réalisés autour du LOM et ses implémentations actuelles et prendre en compte des demandes de modifications exprimées par plusieurs parties prenantes lors des phases de vote et de commentaires. L'enjeu est de trouver un mécanisme de fonctionner de manière compatible avec le LOM tout en le modifiant. La solution a été proposée à l'ISO par l'Afnor (France) pour mettre le LOM au même niveau que ses profils d'application et le situer au niveau « représentation » de la norme ISO 11179. Pour Bourda et Delestre : « Il faut donc définir un modèle conceptuel qui permette d'engendrer tous les schémas pédagogiques, aussi bien le LOM que ses profils d'applications ou que d'autres schémas. C'est ce modèle conceptuel qui constituera le MLR » (Bourda et Delestre, 2005, p. 169).

Parallèlement, l'AUF a contribué de manière significative à l'élaboration de la norme ISO/IEC 2382-36, portant sur le vocabulaire spécifique aux technologies éducatives. Dès sa première version publiée en 2008, cette norme s'est distinguée par son caractère bilingue (anglais et français), soulignant la volonté de faciliter les échanges et la compréhension au sein de la communauté éducative mondiale. L'AUF et le Canada ont veillé à la qualité et à l'enrichissement continu de cette norme lors de ses mises à jour périodiques en 2013 et 2019, et ont proposé des élargissements linguistiques ambitieux impliquant les langues russe, chinoise, coréenne et japonaise. Des initiatives pour intégrer d'autres versions linguistiques comme l'arabe, le vietnamien ou le berbère n'ont cependant pas abouti par manque de soutien officiel de la part de pays concernés. Ces initiatives témoignent toutefois de l'engagement de l'AUF pour une diversité linguistique et culturelle inclusive.

L'AUF a également joué un rôle important dans le choix méthodologique du processus de normalisation terminologique (Ben Henda & Hudrisier, 2009). Face à des longs débats entre les adeptes de la démarche onomasiologique (partant du concept vers le terme) et de la démarche sémasiologique (partant du terme vers le concept), l'AUF a manœuvré par des alliances stratégiques pour trancher la question. Elle a poussé, avec le soutien plusieurs pays membres, vers l'adoption d'une approche sémasiologique qui fait que les concepts soient définis avant les termes qui les représentent dans une langue ou culture cible. L'exemple qui illustre cette approche est la variation terminologique dans la désignation des niveaux scolaires au sein des systèmes éducatifs francophones. Alors qu'en France, la « sixième », par exemple, correspond à la première année de collège, dans d'autres pays francophones, en Tunisie par exemple, le même terme désigne la dernière année du même cycle. L'accord sur la définition du concept permet à chaque pays d'utiliser le terme qui convient à sa langue et à sa culture même si ces termes sont diamétralement opposés.

L'avenir de la norme terminologique ISO/IEC 2382-36 s'oriente vers la création d'une ontologie des technologies éducatives, une proposition soutenue par l'AUF et le Canada. Cette évolution vise à renforcer l'interopérabilité sémantique en structurant un réseau global de connaissances, garantissant la cohérence des contenus numériques éducatifs. L'objectif est de bâtir un cadre permettant un partage plus harmonisé des ressources pédagogiques et une meilleure intégration des divers systèmes éducatifs.

En conclusion, la Francophonie, à travers ses contributions aux travaux du SC36, soutient l'élaboration de normes techniques comme leviers essentiels pour l'harmonisation des pratiques éducatives. Ces actions, basées sur une approche inclusive et respectueuse de la diversité culturelle et linguistique, répondent aux défis de l'interopérabilité et de l'adaptabilité dans le domaine des technologies éducatives, tout en promouvant une représentativité multilingue et une appropriation culturelle accrue des ressources.

### Un regard critique

La normalisation offre de nombreux avantages en favorisant la cohésion, l'efficacité et l'évolutivité dans la société. Elle simplifie les systèmes complexes et assure l'interopérabilité, éléments clés de la vie moderne. Toutefois, elle présente aussi des inconvénients, notamment un manque de flexibilité, des coûts élevés et des freins à l'innovation. Pour que la normalisation reste bénéfique, il est essentiel de trouver un équilibre entre standardisation et adaptabilité, afin de répondre aux besoins sociétaux tout en préservant la diversité et en stimulant l'innovation.

Dans le domaine éducatif, les normes techniques favorisent l'harmonisation des pratiques et l'uniformité des processus d'apprentissage en ligne, même dans un contexte marqué par une grande diversité culturelle et linguistique. Cependant, en l'absence de caractère contraignant, ces normes reposent principalement sur le consensus volontaire où la résistance est forte. Contrairement aux industries où les standards garantissent l'efficacité et la compatibilité, l'éducation reste plus réticente à adopter des normes et des standards, en raison de l'hétérogénéité des approches pédagogiques et des systèmes numériques souvent fermés. Les tentatives de standardisation, telles que le *Learning Design* (IMS-LD) et les métadonnées pédagogiques (LOM), ont eu un succès limité, en grande partie à cause de leur faible adoption par les institutions et les particuliers.

Les critiques adressées à la normalisation sont similaires à celles dirigées contre les régulations en général. Certains considèrent les normes comme une intrusion dans un monde où les individus et organisations devraient pouvoir décider par eux-mêmes. Bien qu'elles apportent de l'ordre, les normes peuvent aussi freiner l'innovation, surtout lorsqu'elles sont trop rigides. Dans le domaine de l'éducation, une sur-standardisation pourrait limiter la pensée critique, l'innovation, et réduire la pertinence des curriculums. De même, dans le domaine de la santé, des protocoles standardisés trop stricts peuvent empêcher des soins adaptés, obligeant les praticiens à suivre des directives qui ne répondent pas toujours aux besoins spécifiques des patients.

En somme, bien que la normalisation soit essentielle pour assurer l'efficacité et l'interopérabilité, il est crucial de maintenir un juste équilibre et de veiller à ce qu'une approche flexible et contextuelle soit privilégiée pour permettre des adaptations nécessaires aux situations particulières et préserver la créativité et la diversité.

## Bibliographie

**Résumé :** Les normes sont adoptées dans un large éventail de domaines aussi bien techniques et industriels que socioéconomiques, culturels et linguistiques. Elles se présentent de manière explicite, telles que les lois, les règlements, les normes techniques et industrielles, ou implicites sous la forme de normes sociales non écrites. Or, dans une mondialisation marquée par une très fine mosaïque d'identités socioculturelles, la question se pose par rapport à la construction de systèmes globaux, transparents et cohérents dans lesquels un travail considérable de consensus est nécessaire pour assurer tous les types de transferts et leurs adaptations locales. La focale est mise ici sur l'écosystème mondial de l'éducation qui développe ses propres normes de transfert de connaissances et de valeurs socioculturelles par l'apprentissage, l'enseignement et la formation. Le Sous-comité 36 de l'organisation mondiale de la normalisation est l'une des structures de cet écosystème auquel participe la Francophonie pour élaborer des normes internationales de l'enseignement à distance sur la base de consensus universels.

**Mots-clés :** Normes et standards, technologies éducatives, Francophonie, Transfert de connaissance

**Abstract :** *Standards are adopted in a wide range of fields, both technical and industrial, as well as socio-economic, cultural and linguistic. They are presented explicitly as laws and regulations, technical and industrial standards or implicitly in the form of unwritten social standards. However, in a globalization marked by a very fine mosaic of socio-cultural identities, the question arises in relation to the construction of global, transparent and coherent systems in which considerable work of consensus is necessary to ensure all types of transfers and their local adaptations. The focus here is on the global education ecosystem which develops its own standards for the transfer of knowledge and socio-cultural values through learning, teaching and training. Subcommittee 36 of the International Organization for Standardization is one of the structures of this ecosystem in which the Francophonie participates to develop international standards for distance education on the basis of universal consensus.*

**Keywords** *: Norms and standards, learning technologies, Francophonie, Knowledge transfer*